# Tuple Value Based Multiplicative Data Perturbation Approach To Preserve Privacy In Data Stream Mining

Hitesh Chhinkaniwala[1] and Sanjay Garg[2]

[1]Department of Computer Engineering, Ganpat University, Mehsana, India
hitesh.chhinkaniwala@ganpatuniversity.ac.in
[2] Department of Computer Engineering, Nirma University, Ahmedabad, India
sgarg@nirmauni.ac.in

## ABSTRACT

*Huge volume of data from domain specific applications such as medical, financial, library, telephone, shopping records and individual are regularly generated. Sharing of these data is proved to be beneficial for data mining application. On one hand such data is an important asset to business decision making by analyzing it. On the other hand data privacy concerns may prevent data owners from sharing information for data analysis. In order to share data while preserving privacy, data owner must come up with a solution which achieves the dual goal of privacy preservation as well as an accuracy of data mining task – clustering and classification. An efficient and effective approach has been proposed that aims to protect privacy of sensitive information and obtaining data clustering with minimum information loss.*

## 1. INTRODUCTION

The data stream paradigm has recently emerged in response to the continuous data problem in data mining. Mining data streams is concerned with extracting knowledge structures represented in models and patterns in non-stopping, continuous streams (flow) of information. Potentially infinite volumes of data streams are often generated by Real-time surveillance systems, Communication networks, Internet traffic, On-line transaction in financial market or retail industry, Electric power grids, Industry production processes, Remote sensors, and other dynamic environments. These data sets need to be analyzed for identifying trends and patterns, which help us in isolating anomalies and predicting future behaviour. However, data owners or publishers may not be willing to exactly reveal the true values of their data due to various reasons, most notably privacy considerations. Providing security to sensitive data against unauthorized access has been a long term goal for the database security research community. Therefore, in recent years, privacy-preserving data mining has been studied extensively. There exist different techniques for privacy preserving data mining.  Privacy preserving data mining techniques are discussed in [1]. Agrawal and Srikant [2] used the random data perturbation technique. More solutions based on data perturbation can be found in [3][4][5][6]. Some heuristic based solutions are also proposed such as k-anonymity [7] to protect the identity of individual or group of entities through micro data release. Improvements on k-anonymity are found in [8][9]. The work presented in [10][11][12] addressed the data privacy based on secure multiparty computations, management of data stream and related works especially clustering and classification of data stream mining are discussed in [13]. The work presented here addresses the issue of data privacy.

The paper is organized with next section presents an overview of framework with proposal of tuple value based data perturbation approach. Section 3 analyse results with standard evaluation parameters. The last section concludes with summary of work and remarks on results.

## 2. RELATED WORK

A number of recently proposed techniques address the issue of privacy preservation by perturbing the data and reconstructing the distributions at an aggregate level in order to perform the mining. The work presented in [2] addresses the problem of building a decision tree classifier in which the values of individual records have been perturbed using randomization method. While it is not possible to accurately estimate original values in individual data records, the authors propose a reconstruction procedure to accurately estimate the distribution of original data values.

The work presented in [3] proposes an improvement over the Bayesian-based reconstruction procedure by using an Expectation Maximization (EM) algorithm for distribution reconstruction. More specifically, the authors prove that the EM algorithm converges to the maximum likelihood estimate of the original distribution based on the perturbed data. Evfiemievski et al. [4] proposed a select-a-size randomization technique for privacy preserving mining of association rules. Du et al. [14] suggested randomized response techniques for PPDM and constructed decision trees from randomized data. Other reconstruction based works discussed in [15][16][17]. Research carried out so far shows the distribution of random noises, recovering the distribution of the original data is possible. Kargupta et al. [18] challenged the additive noise schemes, and pointed out that additive noise might not be secure. They proposed a random matrix-based spectral filtering technique to recover the original data from the perturbed data. Their results have shown that the recovered data can be reasonably close to the original data. The results indicate that for certain types of data, additive noise might not preserve privacy as much as we require for PPDM. Huang and Du [19] excused the privacy breaches using correlation among attributes. They proposed two data reconstruction methods that are based on data correlations. One method uses the Principal Component Analysis (PCA) and the other method uses Bayes estimate technique. They conducted theoretical and experimental analysis on the relationship between data correlations and the amount of private information that can be disclosed based on the data reconstruction method based on PCA and Bayes estimation technique. Their studies have shown that when the correlation between the attributes is high, the original data can be reconstructed more accurately, i.e. more privacy breaches will happen.

Noise addition or multiplication is not the only technique which can be used to perturb the data. A related method is that of data swapping, in which the values across different records are swapped in order to perform the privacy preservation [20]. One advantage of this technique is that the lower order marginal totals of the data are completely preserved and are not perturbed at all. Therefore certain kinds of aggregate computations can be exactly performed without violating the privacy of the data.

## 3. PRELIMINARIES

### 3.1. Framework

To test extended framework as shown in Figure 1, Massive Online Analysis (MOA) has been used. MOA is a software environment for implementing algorithms and running experiments for online learning from evolving data streams [21][22]. MOA supports evaluation of data stream learning algorithms on large streams for both Clustering and Classification. In addition to this it also supports interface with WEKA machine learning algorithms. Following are the steps of using MOA framework,

- Configure Data Stream Source. Let D is Data stream Source.
- Use D as input to MOA Stream learning clustering algorithms.
- Result (R): (1) Evaluation measures for stream clustering (2) Visualization for analyzing result.

We have extended original Massive Online Analysis (*MOA*) framework as shown in figure 1. Dataset $D$ is given as an input to proposed data perturbation algorithm. Algorithm perturbs only sensitive attribute values and resultant dataset with modified values is called perturbed dataset $D'$. $D$ and $D'$ are provided to standard clustering stream learning algorithms to obtain results $R$ and $R'$ respectively. Proposed work focuses on obtaining close approximation between clustering results $R$ and $R'$ to balance trade off between privacy gain and information loss. The primary objective of the second stage, which is handled by the online data mining system, is to mine perturbed data streams to cluster the data.

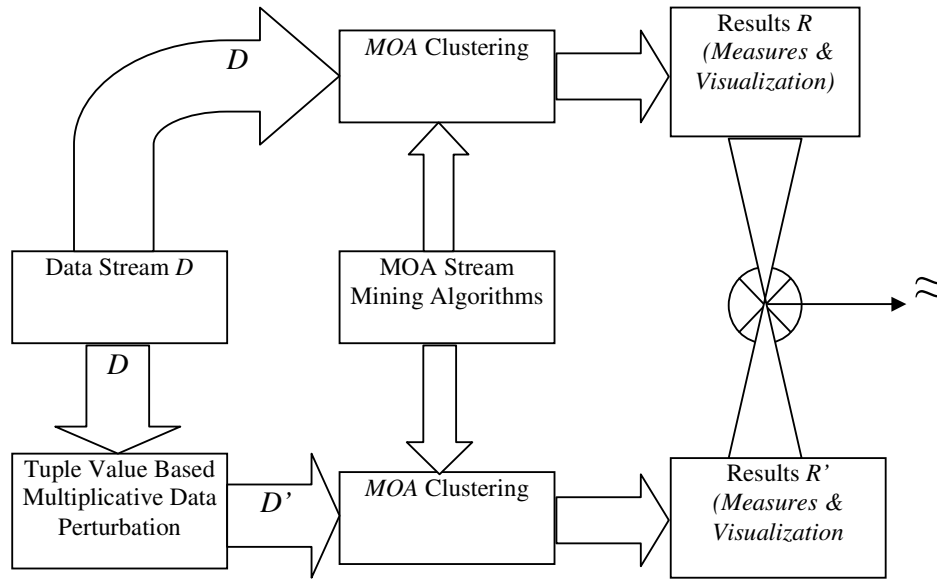

Figure 1. Extended Massive Online Analysis (MOA) framework

## 3.2. Tuple Value Based Multiplicative Data Perturbation

To protect the sensitive attribute value, tuple value of instance to be processed is computed first. Tuple value is the average of normalized values (computed using $Z\text{-}score$ normalization) of attributes of given instance except the class attribute. The tuple values are then multiply with the values of sensitive attribute of respective instances. The resultant dataset with perturbed sensitive attribute values is likely preserves statistical characteristics of original dataset.

## 3.3. Algorithm

**Algorithm**: *Tuple value based multiplicative data perturbation*
**Input**: Data streams D
**Intermediate result**: Perturbed data streams D'
**Output**: Clustering results R and R' of data streams D and D'

**Algorithm steps**
For each instance I in dataset D
TupleValue = 0
For each Attribute $A_J$ in Instance I where $J = 1, 2, 3, ..., Total\ Attributes - 1$

*If not normalized $A_j$*

  *V = Stddev $\left(A_j\right)$*

  *M = Mean $\left(A_j\right)$*

  *NormalizedValue $\left(A_j\right) = \frac{(Value(A_j) - M)}{V}$*

 *Else*

  *NormalizedValue $\left(A_j\right) = Value(A_j)$*

 *Endif*

*TupleValue = TupleValue + NormalizedValue $\left(A_j\right)$*

*End for*

*TupleValue = Average (TupleValue)*

*Value $(I_s) = TupleValue \times Value (I_s)$  //$I_s$ = Instance sensitive attribute*

*Cluster $(T)$*

*End for*

***Function Description:***

*Stddev $\left(A_j\right) = Standard\ deviation\ of\ attribute\ A_j$*

*Mean $\left(A_j\right) = Mean\ of\ attribute\ A_j$*

### 3.4. Experiments

To evaluate the effectiveness of proposed privacy preserving method, Experiments have been carried out on Intel Core I3 Processor with 3 GB primary memory on Windows XP system. Simulation has been done in data stream clustering environment. We quantified proposed approach using resultant accuracy of true dataset clustering and perturbed dataset clustering.

The experiments were processed on two different data sets available from the UCI Machine Learning Repository [23], MOA dataset dictionary [24]. *K-Mean* Clustering algorithm using WEKA data mining tool in MOA framework has been simulated to evaluate the accuracy of proposed privacy preserving approach.

## 4. RESULTS

### 4.1. Cluster Membership Matrix (CMM)

Cluster Membership Matrix identifies how closely each cluster in the perturbed dataset matches its corresponding cluster in the original Dataset in table I. Rows represent the clusters in the original dataset, while Columns represent the clusters in the perturbed dataset, and Freq$_{i, j}$ is the number of points in cluster $C_i$ that falls in cluster $C_i$' in the perturbed dataset.

Table 1.  Cluster Membership Matrix

|  | C₁' | C₂' | ...... | Cₙ' |
|---|---|---|---|---|
| **C₁** | Freq $_{1,1}$ | Freq $_{1,2}$ | ...... | Freq $_{1,n}$ |
| **C₂** | Freq $_{2,1}$ | Freq $_{2,2}$ | ...... | Freq $_{2,n}$ |
| : : | ⋮ | ⋮ | ...... | ⋮ |
| **Cₙ** | Freq $_{n,1}$ | Freq $_{n,2}$ | ...... | Freq $_{n,n}$ |

Table 2 shows datasets configuration to determine the effectiveness of our proposed method. Table 3 shows the percentage of accuracy obtained when selected attribute are perturbed using our algorithm in each dataset.

Table 2.  Datasets

| Dataset | Class Domain | # of Instances | Sensitive Attributes |
|---|---|---|---|
| Covertype[16] | {1,2,3,4,5,6,7} | 65k | Elevation, Slope |
| Electric Norm[17] | {Up, Down} | 45k | Nswprice, Nswdemand |

Table 3.  Accuracy Obtained

| Dataset | Attributes Perturbed | % Accuracy |
|---|---|---|
| Covertype[16] | Elevation, | 98.73 % |
|  | Slope | 99.16 % |
| Electric Norm[17] | Nswprice | 99.97 % |
|  | Nswdemand | 74.76 % |

## 4.2. Precision & Recall Measures

Precision and Recall are two important measures to determine the effectiveness and accuracy of the information retrieval system. Results of proposed approach have been quantified using precision and recall measures provided with MOA framework. Accuracy using these two measures is represented using line graph in figure 2 to figure 9. Each graph represents the measure obtained when original data is processed without applying proposed perturbation approach and when data is undergone through proposed approach. *K-Mean* clustering algorithm is used to generate 5 clusters scenario. Precision and Recall measures have been evaluated with sliding window (w) – 3000 tuples.

$$Precision = \frac{\left( \sum_{I=1}^{|C|} f1\_p(I) \right)}{Realclust} \quad \text{................................................ (1)}$$

Where,

$|C| = Number\ of\ clusters$

$$f1\_p(I) = \frac{2 \times Precision\ (I) \times Recall\ (I)}{Precision\ (I) + Recall\ (I)}$$

$$Precision\ (I) = \frac{Max(C[I,J])}{Clustertotalweight\ (i)}$$

$$Recall\ (I) = \frac{Max(C[I,J])}{Classtotalweight\ (I)}$$

$C\ [I,J]$ = Weight of cluster I and class J

$Clustertotalweight\ (I)$ = Returns the total number of instances belong to cluster I

$Classtotalweight\ (I)$ = Total number of instances belong to class J

$$Recall = \frac{\left(\sum_{J=1}^{|C|} f1\_r(J)\right)}{Numclasses} \qquad \dots\dots\dots\dots\dots\dots\dots\dots\dots\dots\dots\dots\dots\dots\dots\dots\dots\ (2)$$

Where,

$|C|$ = Number of clusters

$$f1\_r(J) = Max\left(f(I = 1,2,3 \dots Number\ of\ Clusters, J)\right)$$

$$f(I,J) = \frac{2 \times Precision\ (I) \times Recall\ (I)}{Precision\ (I) + Recall\ (I)}$$

$$Precision\ (I) = \frac{Clusterclassweight\ (I,J)}{Clustertotalweight\ (I)}$$

$$Recall\ (I) = \frac{Clusterclassweight\ (I,J)}{Classtotalweight\ (J)}$$

$Clusterclassweight\ [I,J]$ = Total instances belong to cluster I and class J

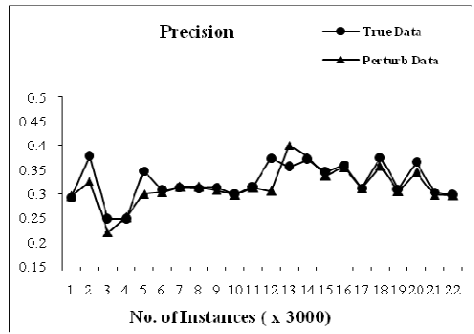

Figure 2. Measured Precision on attribute Elevation in Covertype dataset

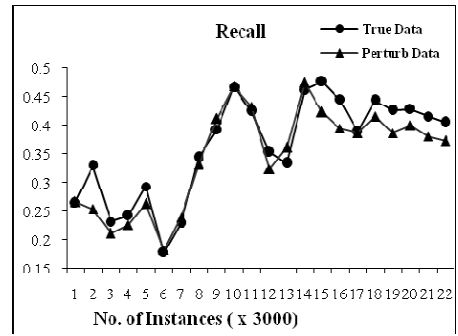

Figure 3. Measured Recall on attribute Elevation in Covertype dataset

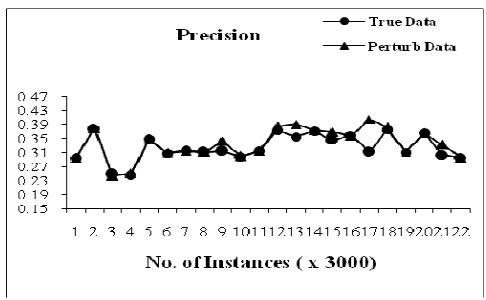

Figure 4. Measured Precision on attribute Slop in Covertype dataset

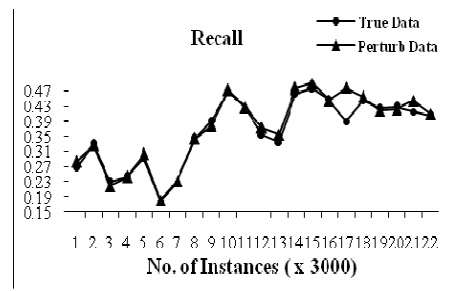

Figure 5. Measured Recall on attribute Slop in Covertype dataset

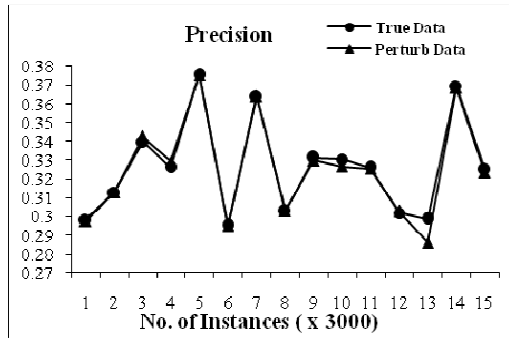

Figure 6. Measured Precision on attribute Nswprice in Electric norm dataset

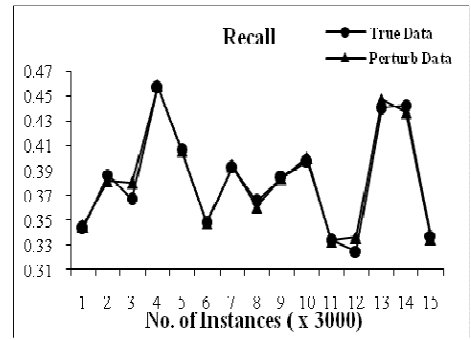

Figure 7. Measured Recall on attribute Nswprice in Electric norm dataset

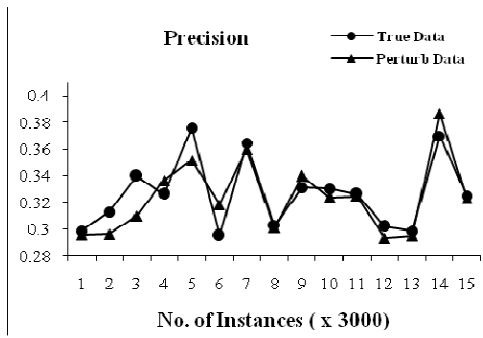

Figure 8. Measured Precision on attribute Nswdemand in Electric norm dataset

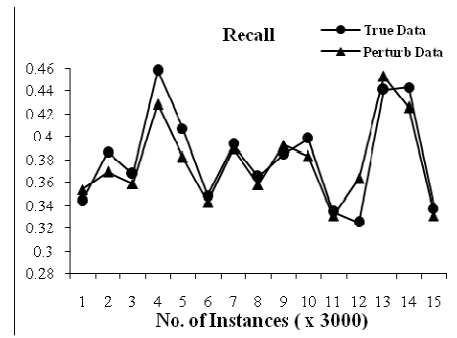

Figure 9. Measured Recall on attribute Nswdemand in Electric norm dataset

## 5. CONCLUSIONS

Proposed approach focused on data perturbation by randomization noise addition to preserve privacy of sensitive attributes. Tuple value based multiplication data perturbation tried to keep statistical relationship among the tuple attributes intact. Proposed approach considered sensitive attribute as dependent attribute and remaining attributes of tuple except class attribute as independent attributes. Independent attributes of tuple has been used to calculate tuple specific random noise. *K-Mean*s clustering algorithm over defined sliding window size on perturbed data stream has been used in order to estimate the accuracy and effectiveness of clustering results over two standard datasets. Results show fairly good level of privacy has been achieved with reasonable accuracy in almost all tested cases. Privacy of original data after applying perturbation has been quantified using misclassification error. Information loss due to data perturbation has been quantified by percentage of instances of data stream that are been misclassified using cluster membership matrix. Proposed approach shows reasonably good results against evaluation measures – Precision, Recall, Misclassification and Cluster Membership Matrix. We limited experiments to protect numeric attributes only but work can be extended to nominal type attributes also.

## REFERENCES

[1] Verykios VS, Bertino K, Fovino IN, Saygin Y & Theodoridis Y, (2004) "State-of-the-Art in Privacy Preserving Data Mining", *ACM SIGMOD*, Record. 33, pp 50-57.

[2] Agrawal R & Srikant R, (2000) "Privacy-preserving data mining", In Proceeding of the *ACM SIGMOD Conference on Management of Data*. ACM Press, pp 439-450.

[3] Agrawal D & Aggarwal C, (2001) "On the design and quantification of privacy preserving data mining algorithms", In Proceedings of the 20th *ACM SIGACT SIGMOD SIGART Symposium on Principles of Database Systems*, pp 247-254.

[4] Evfimievski A, Srikant R, Agrawal R, & Gehrke J, (2002) "Privacy preserving mining of association rules", In Proceedings of the 8th *ACM SIGKDD International Conference on Knowledge Discovery and Data Mining*, pp 217-228.

[5] Rizvi SJ & Haritsa JR, (2002) "Maintaining data privacy in association rule mining", In Proceedings of the 28th *International Conference on Very Large Data Bases*, pp 682-693.

[6] Polat H & Wenliang D, (2003) "Privacy-preserving collaborative filtering using randomized perturbation techniques", In Proceedings of the 3rd *IEEE International Conference on Data Mining,* pp 625-628.

[7] Samarati P, (2001) "Protecting respondents' identities in microdata release", *IEEE transactions on Knowledge and Data Engineering*, pp 1010-1027.

[8] Machanavajjhala A, Gehrke J, Kifer D & Venkitasubramaniam M, (2007) "l-diversity: Privacy beyond k-anonymity", *ACM Transaction on Knowledge Discovery in Data*, Vol. 1, Issue 1, Article No.3.

[9] Li N, Li T & Venkatasubramanian S, (2007) "t-closeness: Privacy beyond k-anonymity and l-diversity", In Proceedings of the 23rd *IEEE International Conference on Data Engineering*, pp 106-115.

[10] Vaidya J & Clifton C, (2003) "Privacy-Preserving K-Means Clustering Over Vertically Partitioned Data", In Proceedings of the 9th *ACM SIGKDD International Conference on Knowledge Discovery and Data Mining*.

[11] Yao AC, (1986) "How to generate and exchange secrets?", In Proceedings of the 27th *IEEE Symposium on Foundations of Computer Science*, pp 162-167.

[12] Clifton C, Kantarcioglu M, Vaidya J, Lin X and Zhu M, (2003) "Tools for Privacy Preserving Distributed Data Mining", In *ACM SIGKDD Explorations*, Vol. 4, No. 2.

[13] Golab L & Ozsu M, (2003) "Issues in data stream management", In *ACM SIGMOD*, Record 32, No. 2, pp 5-14.

[14] W. Du and Z. Zhan, (2003) "Using randomized response techniques for PPDM", In Proceeding of the 9th ACM SIGKDD, pp. 505-510.

[15] K. Liu, H. Kargupta and J. Ryan, (2006) "Random projection-based multiplicative perturbation for privacy preserving distributed data mining", IEEE Transactions on Knowledge and Data Engineering, Vol 18, No. 1, pp. 92-106.

[16] J. Ma and K. Sivakumar, (2005) "Privacy preserving Bayesian network parameter learning", 4th WSEAS International Conference on Computational Intelligence, Man-machine Systems and Cybernetics, Miami, Florida.

[17] J. Ma and K. Sivakumar, (2006) "A PRAM framework for privacy-preserving Bayesian network parameter learning", WSEAS Transactions on Information Science and Applications, Vol 3, No. 1.

[18] H. Kargupta, S. Datta, Q. Wang and K. Sivakumar, (2003) "On the privacy preserving properties of random data perturbation techniques", In Proceeding of the IEEE International Conference on Data Mining, Melbourne, FL. November, pp: 99-106.

[19] Z. Huang, W. Du and B. Chen, (2005) "Deriving private information from randomized data", In Proceeding of SIGMOD, pp. 37-48, Baltimore, Maryland, USA.

[20] S. Fienberg and J. McIntyre, (2003) "Data Swapping: Variations on a Theme by Dalenius and Reiss", Technical Report, National Institute of Statistical Sciences.

[21] Bifet A, Kirkby R, Kranen P & Reutemann P, (2011) "Massive Online Analysis Manual".

[22] Bifet A, Holmes G & Pfahringer B, (2010) "Massive Online Analysis, a Framework for Stream Classification and Clustering", In Proceedings of the *JMLR Workshop and Conference*, pp 44-50.

[23] UCI Data Repository, http://archive.ics.uci.edu/ml/datasets (accessed July 2012).

[24] MOA datasets, http://moa.cs.waikato.ac.nz/datasets (accessed June 2012).

## Authors

Hitesh Chhinkaniwala is Ph.D scholar of KSV University, Gujarat, India. He has done B.E. in year 1997 and M.Tech in Computer Science & Engineering in year 2008 from NIT Karnataka. Currently working as an assistant professor at U.V. Patel College of Engineering, Ganpat University, Gujarat, India. He has 13 years of teaching experience. He has carried out his PG- 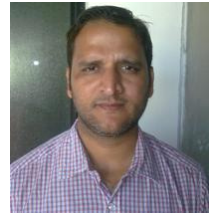
Dissertation on Inter-transaction Association Rule Mining. His areas of interest are data mining, data & information security, Privacy preserving data mining and statistical analysis.

Dr Sanjay Garg has an experience of more than 20 years in the field of academics including experience at SATI Vidisha(M.P) and at ADIT Vallabh Vidhya Nagar(Gujarat). He has published numbers of papers in the area of Data Mining, Pattern Recognition and Data Base Systems at International and National level. Dr Garg is a senior member IEEE and member of ACM. He has guided 20 PG dissertations and currently guiding six Ph.D. scholars. 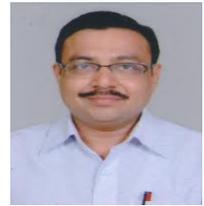